\documentclass[12pt]{iopart}
\usepackage{iopams}
\usepackage{graphicx}
\newcommand{\cms}{Co$_2$MnSi}
\newcommand{\cfs}{Co$_2$FeSi}
\begin{document}
\title{Hall effect in laser ablated Co$_2$(Mn,Fe)Si thin films}

\author{H Schneider, E Vilanova}
\address{Institut f\"ur Physik, Johannes Gutenberg-Universit\"at, 55099 Mainz, Germany}
\author{B Balke, C Felser}
\address{Institut f\"ur Anorganische Chemie und Analytische Chemie, Johannes Gutenberg-Universit\"at, 55099 Mainz, Germany}
\author{G Jakob}
\address{Institut f\"ur Physik, Johannes Gutenberg-Universit\"at, 55099 Mainz, Germany}

\ead{schneiho@uni-mainz.de}
\begin{abstract}
Pulsed laser deposition was employed to grow thin films of the Heusler compounds \cms{} and \cfs{}. Epitaxial growth was realized both directly on MgO (100) and on a Cr or Fe buffer layer. Structural analysis by x-ray and electron diffraction shows for both materials the ordered L2$_1$ structure. Bulk magnetization was determined with a SQUID magnetometer. The values agree with the Slater-Pauling rule for half-metallic Heusler compounds. On the films grown directly on the substrate measurements of the Hall effect have been performed. The normal Hall effect is nearly temperature independent and points towards a compensated Fermi surface. The anomalous contribution is found to be dominated by skew scattering. A remarkable sign change of both normal and anomalous Hall coefficients is observed on changing the valence electron count from 29 (Mn) to 30 (Fe).
\end{abstract}

\submitto{\JPD}
\maketitle


\section{Introduction}
Due to their expected half-metallicity the use of Heusler alloys has been considered for a number of spintronic applications \cite{YAK06,PAL03,SAK05}. Especially the fabrication of magnetic tunneling junctions with Heusler electrodes has advanced rapidly over the last years. Recent experiments with \cms{} electrodes and an amorphous AlO$_x$ barrier suggest a spin polarization of nearly 90\,\% for the relevant layers at the interface between electrode and barrier \cite{SAK05}. However, this value is reached only at temperatures below 4\,K. At room temperature these junctions show a transport spin polarization comparable to elements with conventional ferromagnetic electrodes, despite a Curie temperature of 985\,K. It has been proposed that this decrease in spin polarization is due to the unfortunate position of the Fermi energy near the edge of the half-metallic gap \cite{SAK05,BAL06}. The calculations in \cite{BAL06} further predict that by replacing Mn with Fe the additional electrons will cause a shift of $E_F$ across the half-metallic band gap, while the size of the band gap as well as the general structure of the bands near $E_F$ will stay nearly unchanged. For \cfs{} the Fermi level is then expected to be located below the conduction band of the minority electrons.

The complexity of tunneling junction experiments is likely to introduce additional, barrier dependent contributions to the tunneling currents which will obfuscate intrinsic material properties \cite{MOO95,SON05}. In contrast the measurement of the Hall effect in ferromagnets is a well-established and comparatively simple procedure, which has already been applied to various Heusler alloys \cite{HOR00, HUS06, GOF07}. Although it is not possible to directly map the band structure or measure spin polarization, the electron doping in the Co$_2$(Mn,Fe)Si system is expected to cause changes in the normal as well as in the anomalous regime.

Since the Hall voltage scales with the inverse thickness of the measured sample, it is very advantageous to possess high quality thin films. In this paper we report pulsed laser deposition of \cms{} and \cfs{} on MgO (100) substrates. In contrast to the more widely used sputter deposition no buffer gas is needed and UHV conditions can be sustained during the whole preparation process. The ability of this technique to grow thin Heusler films has been demonstrated by Wang \emph{et al} \cite{WAN05}. We will discuss the effect of Cr and Fe buffer layers on film growth and magnetic properties. It is also possible to achieve epitaxial growth directly on the insulating substrate. Using a standard photolithographic procedure the latter set of films was patterned with a Hall bar structure. The results from these experiments will be discussed in the second part of this paper.

\section{Film growth}
\begin{figure}
\includegraphics[width=\columnwidth]{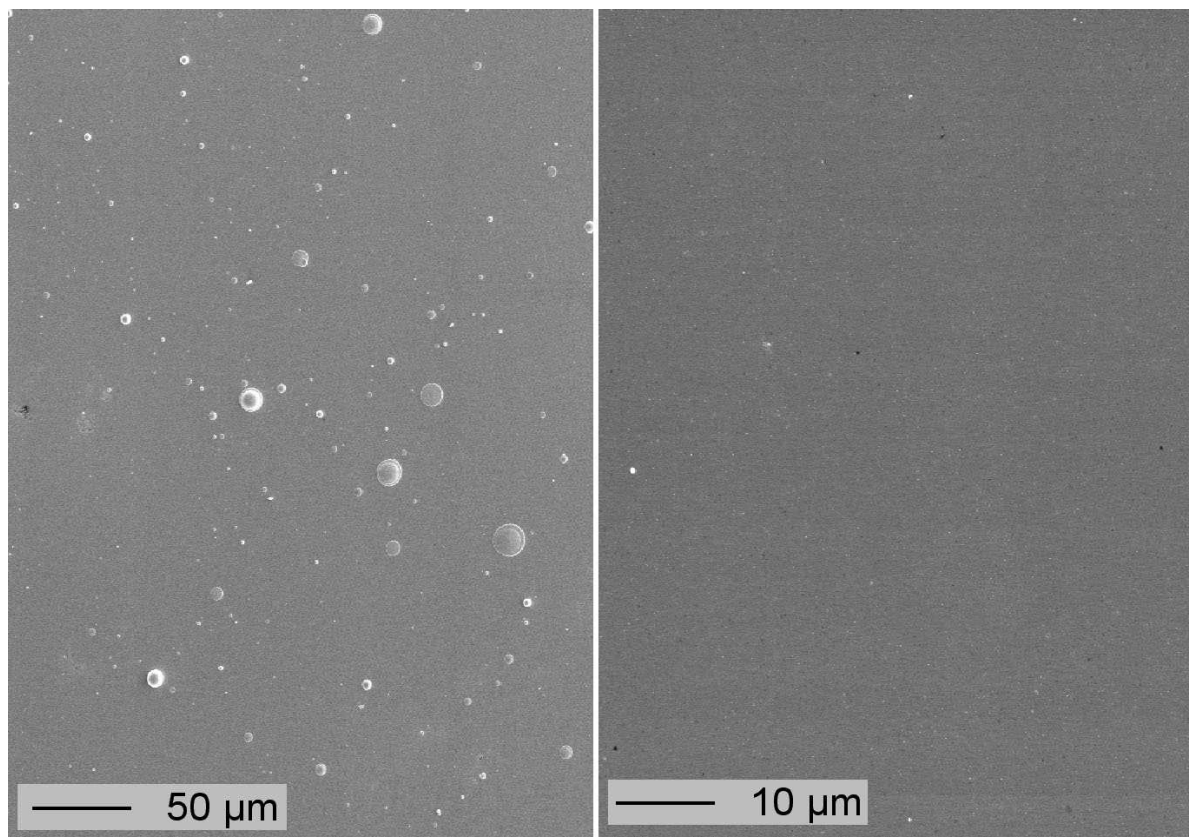}
\caption{SEM images of films deposited with a pulse energy of 700 (left) and 300 mJ (right).}\label{droplets}
\end{figure}

Film growth was realized by pulsed laser deposition (PLD) from stoichiometric targets. As light source a KrF excimer laser ($\lambda=248\,\mathrm{nm}$) with pulse energies ranging from 300 to 700\,mJ was used. The choice affects the growth rate, which increases from 0.2\,\AA{}/min to 2\,\AA{}/min. Deposition was carried out in a UHV chamber ($p_\mathrm{base}\lesssim 2\cdot 10^{-10}\,\mathrm{mbar}$). A RHEED system allowed \emph{in situ} surface analysis. Since no buffer gas was used in the growth process, the formation of melt droplets is observed (see figure \ref{droplets}). For high pulse energies the volume of these droplets is comparable to the film volume. This prohibits the determination of bulk magnetic properties in these films. The situation is drastically improved with lower laser energies. At 300 mJ only a low number of droplets is visible. They contribute to less than 5\,\% of the sample volume. The films were deposited on MgO (100) substrates. In order to overcome the relatively large lattice mismatch of 5.6\,\%, two strategies were employed. One set of samples was deposited directly on the substrate. In this case high substrate temperatures are required to grow the compounds epitaxially. Alternatively the Heusler alloys were deposited onto an intermediate Fe or Cr buffer layer at room temperature and subsequently annealed at 400 -- 500\,$^\circ$C. The buffer layers were also grown by PLD in the same chamber. In order to protect the films from oxidation they were covered with an Al capping layer before transferring them out of the vacuum system.

\section{Structural and magnetic properties}
\begin{figure}
\includegraphics[width=\columnwidth]{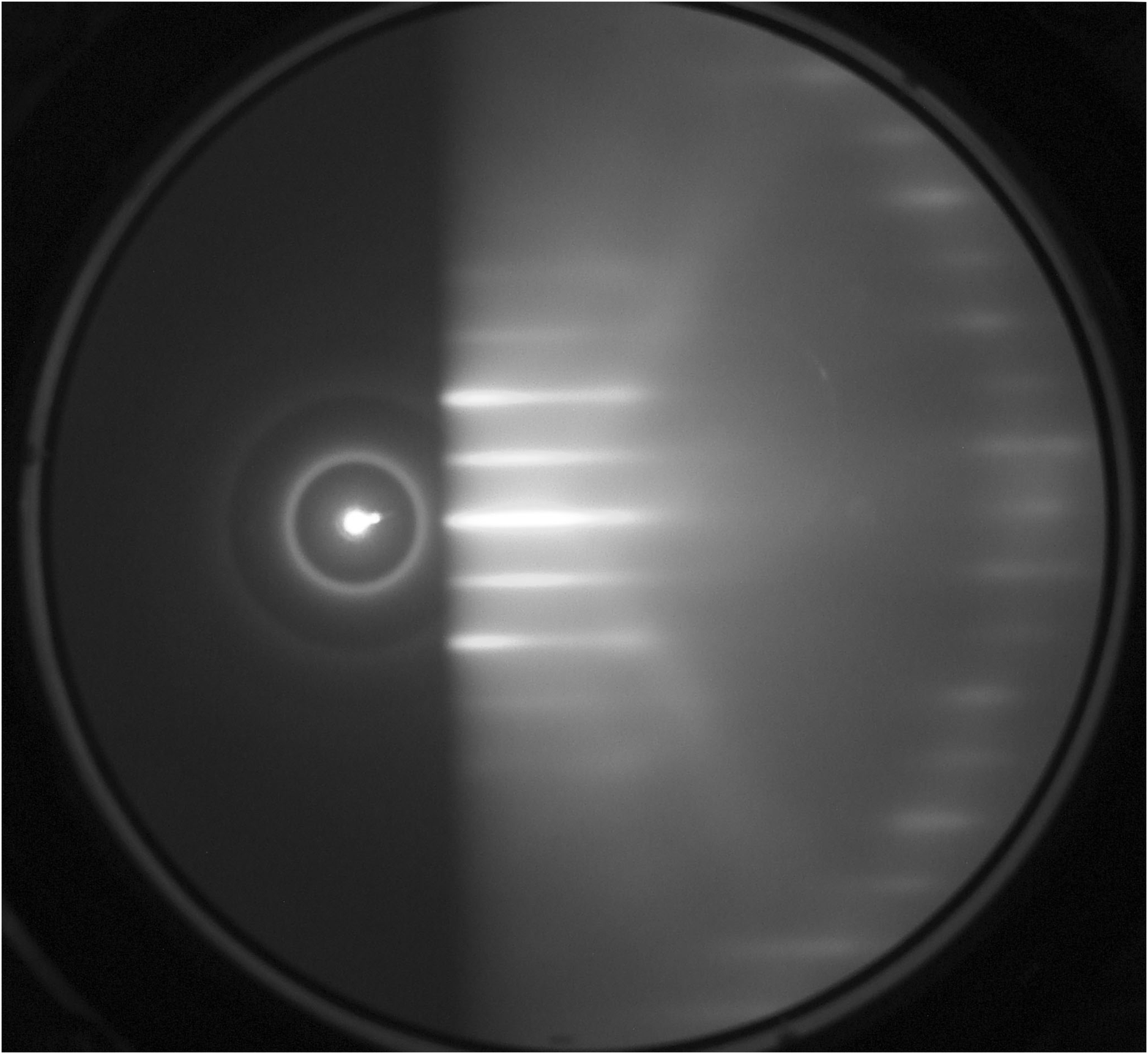}
\caption{RHEED pattern of a \cms{}/Fe/MgO film after annealing at 450\,$^\circ$C.}\label{rheed}
\end{figure}

\begin{figure}
\includegraphics[angle=270,width=\columnwidth]{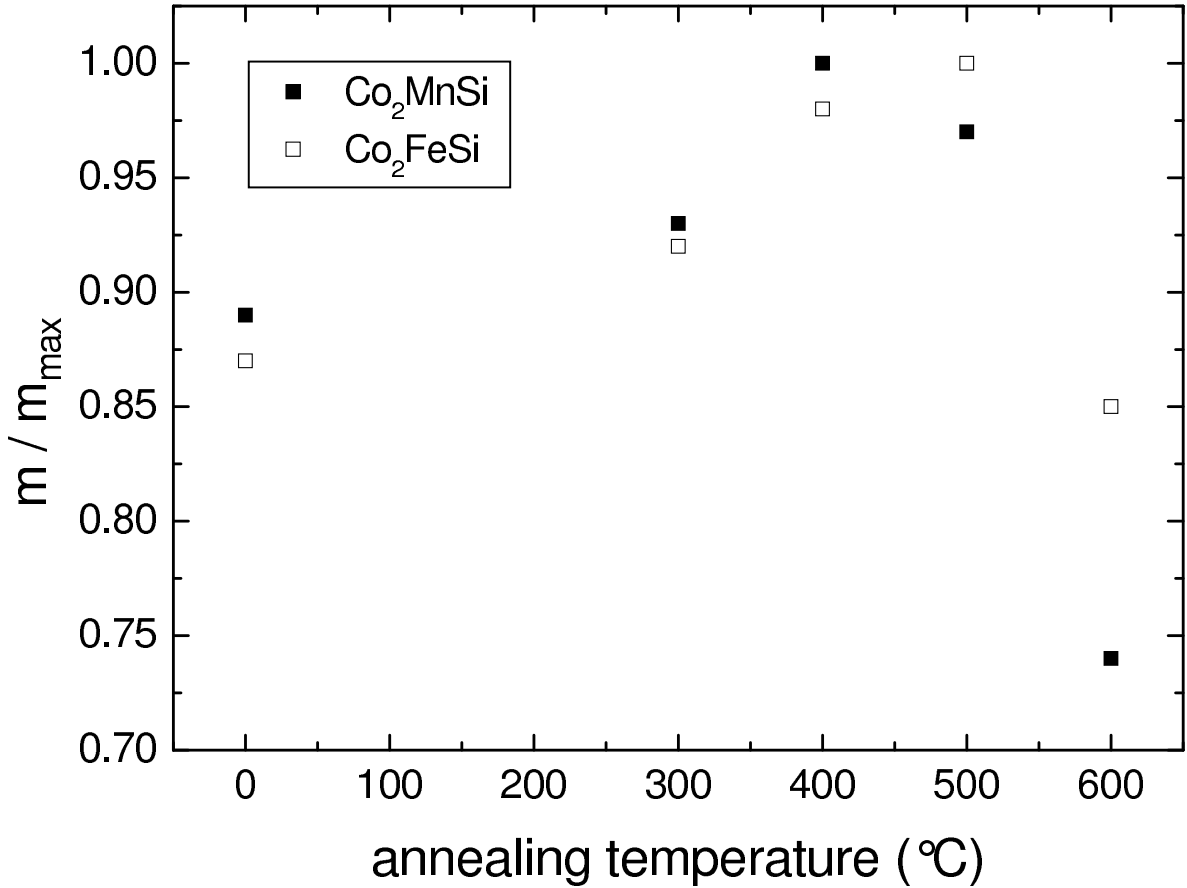}
\caption{Dependence of the sample magnetization on annealing temperature. The films measured here were deposited on a Cr buffer layer.}\label{anneal}
\end{figure}

\begin{figure}
\includegraphics[angle=270,width=\columnwidth]{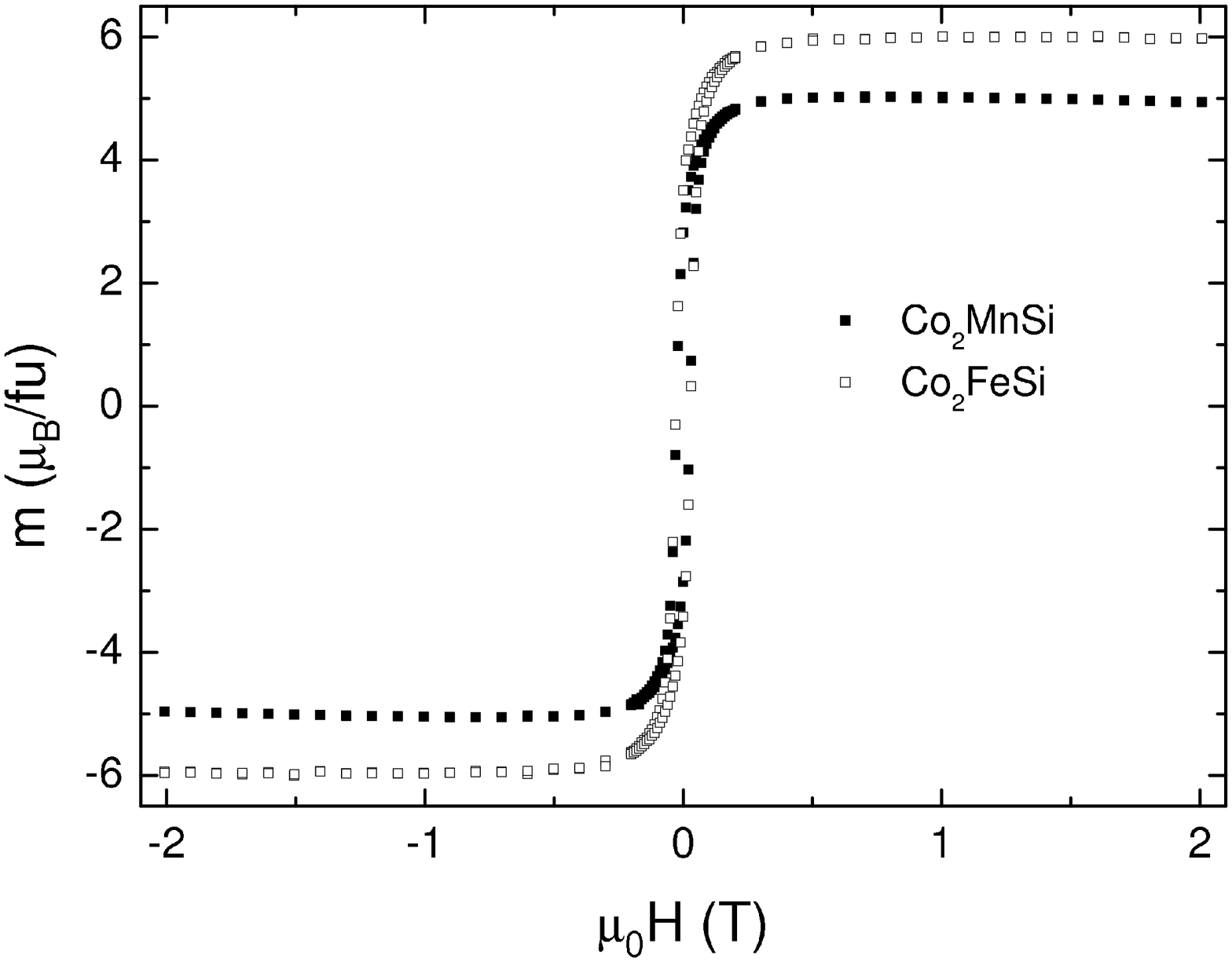}
\caption{Hysteresis loops for Heusler films with Cr buffer layer after annealing at 400\,$^\circ$C (\cms{}) and 500\,$^\circ$C (\cfs{}). Data was taken at a temperature of 20\,K and the background from the substrates was substracted. Because of the unknown contribution of the melt droplets shown in figure \ref{droplets} a relative error of 5\,\% has to be taken into account.}\label{squid}
\end{figure}

The crystal structure of the Heusler films was analyzed by x-ray and electron diffraction. RHEED images as presented in figure \ref{rheed} reveal a smooth and ordered surface. For the films grown on a Cr or Fe buffer $\omega$-$2\theta$ scans of the specular as well as $\phi$ scans of the off-specular reflections show epitaxial, single-phase growth already after deposition. However, the \{111\} reflections cannot be detected, indicating growth not in the fully ordered L2$_1$ but in the B2 structure. After annealing above 300\,$^\circ$C the superstructure reflections can be detected. For an annealing time of 60 minutes the maximum relative intensity is found between 400 and 500\,$^\circ$C. Above 500\,$^\circ$C the scattered intensity of all reflections starts to diminish. Comparison with reports for other Heusler films suggest that this decrease is caused by interdiffusion with the buffer layer \cite{TEZ06, JOU07}.

If the film is deposited directly onto the substrate at room temperature, no long range crystallographic ordering is observed. For these films, a post-annealing procedure does not induce crystalline growth either. Instead substrate temperatures of 400\,$^\circ$C and above are required during deposition. Like the films deposited on a metallic buffer layer, these samples exhibit (100)-oriented and L2$_1$ ordered growth. This is in accordance with sputtered \cfs{} films we reported earlier \cite{SCH06,SCH07}, but in contrast to reports from Inomata \emph{et al} who found long-range crystallographic ordering already at room temperature \cite{INO06}.

Bulk magnetic properties were analyzed with a SQUID magnetometer. As shown in figure \ref{anneal} the film magnetization is correlated with their crystalline quality: For as-deposited films on a metallic buffer layer, the saturation magnetization of both compounds is reduced by about 10 -- 15\,\% compared to the films annealed at 400 -- 500\,$^\circ$C. For these temperatures the magnetization assumes values consistent with the Slater-Pauling rule for Heusler alloys (see figure \ref{squid}) \cite{GAL02}. A further increase of the annealing temperature causes a rapid drop of the magnetization. For the films fabricated without buffer layer the saturation magnetization increases by 25\% from the disordered films to the Slater-Pauling values at substrate temperatures of 600\,$^\circ$C and above. 

\section{Hall effect}

\begin{figure}
\includegraphics[angle=270,width=\columnwidth]{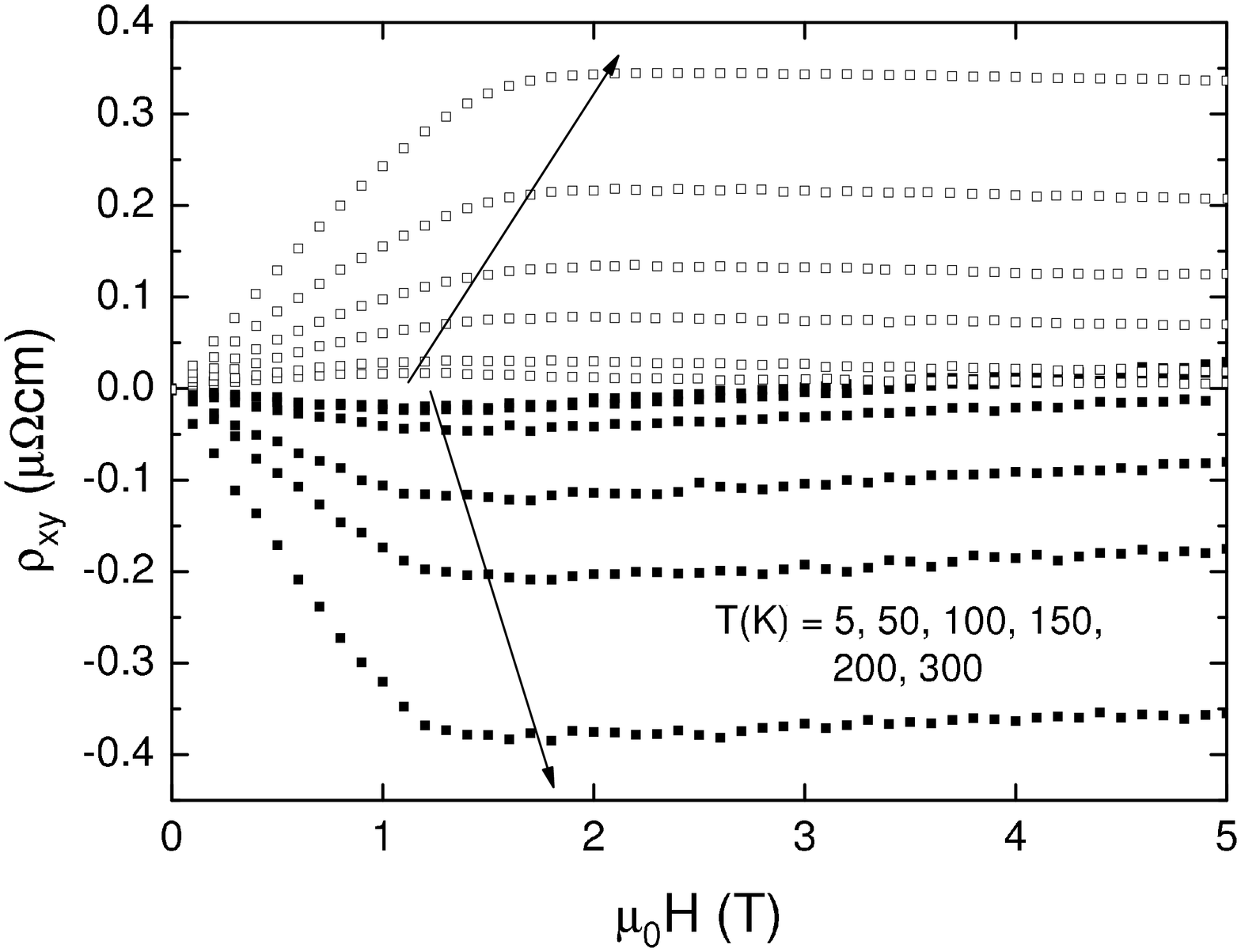}
\caption{Hall resistivity as a function of the applied field. Open symbols represent \cfs{}, closed symbols \cms{}}\label{hall}
\end{figure}

\begin{figure}
\includegraphics[angle=270,width=\columnwidth]{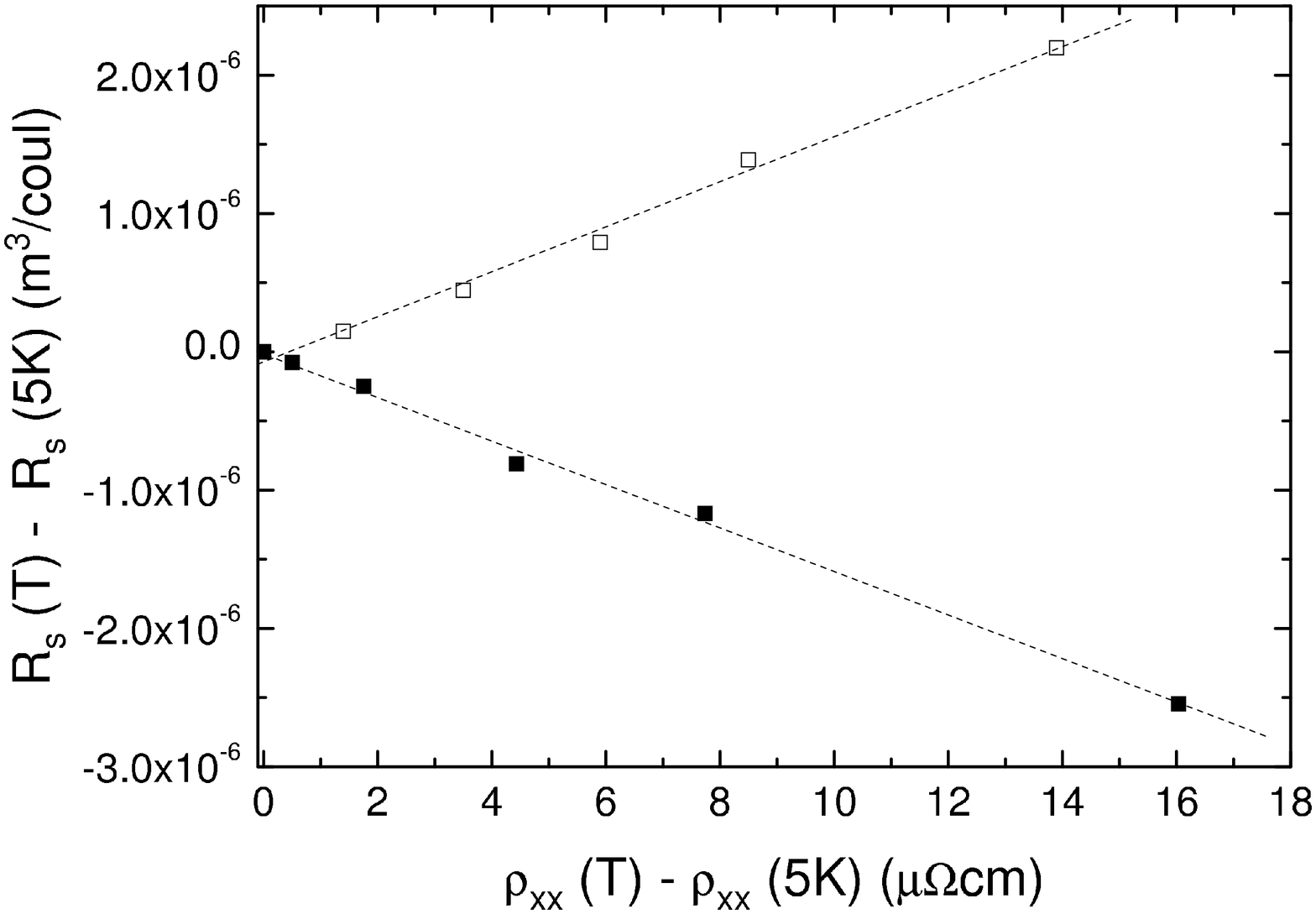}
\caption{Anomalous Hall coefficient as a function of the longitudinal resistivity, normalized to the values at 5\,K. The values for $R_s$ were obtained from figure \ref{hall} by extrapolation of the normal Hall effect to zero field. The dashed lines indicate linear fits. Open symbols represent \cfs{}, closed symbols \cms{}.}\label{skew}
\end{figure}

The availability of epitaxial thin films grown directly on insulating substrates allowed the measurement of the Hall resistivity. For this purpose films were patterned into a Hall bar geometry using a standard photolithographic procedure. A DC current was applied, the measured voltage was antisymmetrized with respect to the external field in order to eliminate ohmic contributions to the signal. To rule out effects from both sample oxidation and a metallic Al cap samples with and without capping layer were compared. These results showed no significant difference.

Figure \ref{hall} shows Hall resistivities for \cms{} and \cfs{} measured at various temperatures. Film thickness in both cases was 50\,nm.  The graphs show the typical behaviour of ferromagnetic materials. Below saturation the signal is governed by the anomalous Hall effect and the increase of the sample magnetization. For higher fields the normal Hall effect caused by the Lorentz force from the external field determines the value of the resistivity. This behaviour is conventionally summed up in the formula $\rho_{xy}(B,M)=R_0B+R_s\mu_0M$ with the normal and anomalous Hall coefficients $R_0$ and $R_s$.

The slope of the normal Hall effect takes a different sign for the two materials. The Fermi surface of Heusler compounds is expected to be multi-sheeted but has not been calculated for the materials in question so far. Therefore a simple Fermi sphere was assumed for a first analysis. In this case \cms{} and \cfs{} appear to have a charge carrier concentration of 5 holes and 25 electrons per formula unit respectively. These high values imply the presence of partially compensated hole-like and electron-like sheets of the Fermi surface. It is noteworthy that in \cfs{} the electron-like contribution seems to dominate although it is closer to a completely filled shell than \cms{}. This indicates that a Fermi sphere oversimplificates the electronic structure of the examined materials.

The anomalous Hall effect in both materials shows also opposite signs -- in the low temperature regime as well as in the temperature dependence. In order to understand the nature of the anomalous Hall effect the anomalous Hall coeffient $R_s$ is plotted as a function of the longitudinal resistivity in figure \ref{skew}. For both materials a linear dependence was found. This implies that the temperature dependence of the anomalous Hall effect is governed by skew scattering. The opposite sign of the slopes is possibly connected to the apparent charge deduced from the normal Hall effect.

\section{Summary}
It has been shown that pulsed laser deposition is a suitable method for the deposition of high quality Heusler alloy thin films: epitaxial growth of \cms{} and \cfs{} on MgO has been realized. The detection of the L2$_1$ crystal structure as well as the adherence of the Slater-Pauling rule for Heusler alloys are consistent with half-metallicity. Since it was possible to omit the use of a buffer layer, electronic transport measurements could be performed. The dependence of the normal Hall coefficient on the electron number in the compounds is counter-intuitive and points towards a  complicated topology of the Fermi surfaces. The evolution of the anomalous Hall coefficient is consistent with skew scattering as the source of the magnetization dependent contribution.

\ack
Financial support by the DFG Research Unit ``New Materials with High Spin Po\-la\-ri\-za\-tion'' (grant no: FOR 559) is gratefully acknowledged.

\end{document}